\documentstyle[12pt]{article}
\newfont{\Mb}{msbm10}

\newcommand{\Od}[2]{{{{\rm d} #1}\over{{\rm d} #2}}}

\begin{document}
\setcounter{equation}{0}
\setcounter{figure}{0}
\setcounter{table}{0}

\title{A Method to Tackle First Order Ordinary 
Differential Equations with Liouvillian Functions in the Solution}
\author{
L.G.S. Duarte\thanks{
Universidade do Estado do Rio de Janeiro,
Instituto de F\'{\i}sica, Departamento de F\'{\i}sica Te\'orica,
R. S\~ao Francisco Xavier, 524, Maracan\~a, CEP 20550--013,
Rio de Janeiro, RJ, Brazil. E-mail: lduarte@dft.if.uerj.br},
S.E.S. Duarte\thanks{
idem. E-mail: sduarte@dft.if.uerj.br}, 
and L.A.C.P. da Mota\thanks{
idem. E-mail: damota@dft.if.uerj.br}
}
\maketitle
\abstract{We present an algorithm to solve First Order Ordinary Differential
Equations (FOODEs) extending the Prelle-Singer (PS) Method. The usual
PS-approach miss many FOODEs presenting Liouvillian functions in the solution (LFOODEs).
We point out why and propose an algorithm to solve a large class of
these previously unsolved LFOODEs. Although our algorithm does not cover all the
LFOODEs, it is an elegant extension mantaining the semi-decision nature of the usual PS-Method.}
\newpage
\section{Introduction}
The problem of solving ordinary differential equations (ODEs) has led,
over the years, to a wide range of different methods for their solution.
Along with the many techniques for calculating tricky integrals, these
often occupy a large part of the mathematics syllabuses of university
courses in applied mathematics round the world. 

The overwhelming majority of these methods 
are based on classification of the DE into types for which a method
of solution is known, which has resulted in a gamut of methods that
deal with specific classes of DEs. This scene changed somewhat at the
end of the 19th century when Sophus Lie developed a general method to
solve (or at least reduce the order of) ordinary differential equations
(ODEs) given their symmetry transformations ~\cite{step,bluman,olver}. Lie's method
is very powerful and highly general, but first requires that we find
the symmetries of the differential equation, which may not be easy to do.
Search methods have been developed~\cite{nosso,nosso2} to extract
the symmetries of a given ODE, however these methods are heuristic and
cannot guarantee that, if symmetries exist, they will be found.
A big step forward in constructing an
algorithm for solving first order ODEs (FOODEs) analytically was taken in a
seminal paper by Prelle and Singer (PS)~\cite{PS} on autonomous systems of
ODEs. Prelle and Singer's problem is equivalent to asking when a FOODE
of the form $y'=M(x,y)/N(x,y)$, with $M$ and $N$ polynomials in their
arguments, has an elementary solution (a solution which can be written
in terms of a combination of polynomials, logarithms, exponentials
and radicals). Prelle and Singer were not exactly able to construct an
algorithm for solving their problem, since they were not able to define
a degree bound for the polynomials which might enter into the
solution. Though this is important from a theoretical point of view, any
pratical use of the PS method will have a degree bound imposed by the 
time necessary to perform the actual calculation needed to handle the
ever-more complex equations. With this in mind it is possible to say
that Prelle and Singer's original method is almost an algorithm, awaiting
a theoretical degree bound to turn it algorithmic. 
In \cite{PS2}, we have presented an extension of the PS-ideas to deal with second order ordinary differential equations (SOODEs). The purpose of this paper is to present another extension to the Prelle-Singer
method allowing for the solution of a class of LFOODEs.
The paper is organized as follows: in section~\ref{PSreview}, we present a short
theoretical introduction to the PS approach; in the following section, we show why the Prelle-Singer aproach misses some LFOODEs; in section~\ref{findingIF}, we
introduce the main ideas of our extension with examples of its application; we
finally present our conclusions.
\section{The Prelle-Singer Procedure}
\label{PSreview}
Despite its usefulness in solving FOODEs, the Prelle-Singer procedure is
not very well known outside mathematical circles, and so we present
a brief overview of the main ideas of the procedure.
Consider the class of FOODEs which can be written as
\begin{equation}
\label{FOODE}
y' = {\frac{dy}{dx}} = {\frac{M(x,y)}{N(x,y)}}
\end{equation}
where $M(x,y)$ and $N(x,y)$ are polynomials with coefficients in the complex 
field $\it C$.
If $R$ is a integrating factor of~(\ref{FOODE}), it is possible to write
$\partial_x (RN) + \partial_y (RM) = 0$, leading to 
$N\, \partial_x R + R\,\partial_y N + M\, \partial_y R + R\, \partial_y M = 0.$ Thus, we finally obtain:
\begin{equation}
\label{eq_PS}
{\frac{D[R]}{R}} = - \left( \partial_x N + \partial_y M\right),
\end{equation}
where $D \equiv N \partial_x + M \partial_y.$
In \cite{PS}, Prelle and Singer showed that, if the solution of (\ref{FOODE}) is written in terms of elementary functions, then $R$ must be of the form $R = \prod_i f^{n_i}_i$ where $f_i$ are irreducible polynomials and $n_i$ are non-zero rational numbers. Using this result in (\ref{eq_PS}), we have
\begin{eqnarray}
\label{ratio}
{\frac{D[R]}{R}} & = & {\frac{D[\prod_{i} f^{n_i}_i]}{\prod_i f^{n_k}_k}} =
{\frac{\sum_i f^{n_i-1}_i n_i D[f_i] \prod_{j \ne i}
f_j^{n_j}}{\prod_k f^{n_k}_k}} \nonumber \\[3mm] 
& = & \sum_i {\frac{f^{n_i-1}_i n_i D[f_i]}{f_i^{n_i}}} =
\sum_i {\frac{n_iD[f_i]}{f_i}}.
\end{eqnarray}
From~(\ref{eq_PS}), plus the fact that $M$ and $N$ are polynomials, 
we conclude that ${D[R]}/{R}$ is a polynomial. One can then prove from~(\ref{ratio}) that $f_i | D[f_i]$ \cite{PS}.
We now have a criterion for choosing the possible $f_i$ (build all
the possible divisors of $D[f_i]$) and, by using~(\ref{eq_PS})
and~(\ref{ratio}), we have
\begin{equation}
\label{eq_ni}
\sum_i {\frac{n_iD[f_i]}{f_i}} = - \left( \partial_x N + \partial_y M\right).
\end{equation}
If we manage to solve~(\ref{eq_ni}) and thereby find $n_i$,
we know the integrating factor for the FOODE and the problem is
reduced to a quadrature.
\section{The PS-Method and LFOODEs}
\label{algo}
In the previous section, the main ideas and concepts used in the Prelle-Singer
method were summarized. In this section, we will first explain why the PS-method
fails to solve some LFOODEs. Then,
using the results on \cite{singer} , we will show that it is possible to
algoritimicaly solve a large class of these.

\subsection{Introduction}
As we have already mentioned, the usual PS-method garantees to find a solution
for a FOODE if it is expressible in terms of elementary functions. However, the
method can also solve some FOODEs with non-elementary solutions. Why and why
not for all the FOODEs with non-elementary solutions? 
Consider the following two examples:
\begin{equation}
\label{Kamke211}
\Od{y}{x}=\frac {3\,{x}^{2}y^{2}+{x}
^{3}+1}{4 \left (x+1\right )\left ({x}^{2}-x+1\right )y}
\end{equation}
and
\begin{equation}
\label{Kamke21}
\Od{y}{x}= y^{2}+y x+x-1,
\end{equation}
equation I.18 of the standard testing 
ground for ODE solvers by Kamke~\cite{kamke}.
These equations present (respectively) general solutions given by:
\begin{equation}
\label{solKamke211}
y^{2}-\sqrt {x+1}\sqrt {{x}^{2}-x+1} \left(
1/2\,\int \!{
\frac {1}{\sqrt {x+1}\sqrt {{x}^{2}-x+1}}}{dx}-2{\it \_c_1} \right) =0,
\end{equation}
and
\begin{equation}
\label{solKamke21}
y =-{\frac {2\,{\it \_C1}+\sqrt {-1}\sqrt {\pi }{e^{-2}}\sqrt {2}{
\it erf}(1/2\,\sqrt {-1}\sqrt {2}x-\sqrt {-1}\sqrt {2})-2\,{e^{1/2\,x
\left (x-4\right )}}}{2\,{\it \_C1}+\sqrt {-1}\sqrt {\pi }{e^{-2}}
\sqrt {2}{\it erf}(1/2\,\sqrt {-1}\sqrt {2}x-\sqrt {-1}\sqrt {2})}}
.
\end{equation}

We note that both solutions are not expressible in terms of elementary
functions. But, for FOODE \ref{Kamke211}, the standard PS-method can find
the solution (eq. \ref{solKamke211}). The same is not true for FOODE
\ref{Kamke21}, what is happening? 
This can be best understood if we have
a look on the integrating factors for those FOODEs, which are respectively:
\begin{equation}
\label{R211}
R = \left ({x}^{3}+1\right )^{-3/2},
\end{equation}
and
\begin{equation}
\label{R21}
R = {\frac {{e^{{x}^{2}/2-2\,x}}}{\left (y+1\right )^{2}}}.
\end{equation}
Since the standard PS procedure constructs integrating factor candidates from
polynomials in the variables ($x,y$), one can see that, since the integrating factor on eq. \ref{R21} presents the exponential
$e^{{x}^{2}/2-2\,x}$, it will never be found by the PS-method. 
So, can we understand something about the general structure of the FOODE that
will allow us, eventually, to solve a class of these latter equations?
\subsection{Important results concerning the general form of the integrating factor for LFOODEs}
In this section, we are going to present a result concerning the general form
for the integrating factor for LFOODEs. Based on that, we are going to
develop, on the next section, a method of solving a class of those LFOODEs.
Using the following theorem, which has been presented on \cite{singer}, 
\bigskip 
{\bf Theorem 1:} {\it Consider a LFOODE of the form $dy/dx=M(x,y)/N(x,y)$, where $M(x,y)$
and $N(x,y)$ are polynomials in $(x,y)$. Its integrating factor must be of the form
$e^{(\int(Udx+Vdy))}$, where $U$ and $V$ are rational functions on $(x,y)$ with
$U_y = V_x$.}
\bigskip
we can extract following result:
\bigskip
{\bf Theorem 2:} Consider a LFOODE of the form $dy/dx=M(x,y)/N(x,y)$, where $M(x,y)$
and $N(x,y)$ are polynomials in $(x,y)$. Its integrating factor must be of the form: 
\begin{equation} 
R = e^{r_0(x,y)} \prod_{i=1}^{n} p_i(x,y)^{c_i}.
\end{equation} 
where $r_0$ is a rational function on $(x,y)$, the $p_i$'s are irreducible polynomials on $(x,y)$
and the $c_i$'s are constants. 
\bigskip

{\bf Proof:} From well known results concerning formal integration \cite{davenport}, we
have that if $f(x)$ is a rational function of $x$, we can
conclude that:
\begin{equation}
\label{IntFx}
\int\!f(x)\,{dx}=r_0(x)+\sum_{i=1}^{K} C_i\, \ln(r_i(x))
\end{equation}
\noindent
where $r_i \,\, (i=0,...,K)$ are rational functions of $x$, and $C_i \,\,
(i=1,...,K)$ are constants.
Let $U(x,y)$ and $V(x,y)$ be two functions such that $U_y = V_x$. So, we can
choose a function $\omega(x,y)$ such that $d\omega = U_x \, dx + V_y \, dy$,
i.e., we can write $\,\omega_x=U$ and $\,\omega_y=V$. Integrating this system (to
obtain $\omega$) we have
\begin{equation}
\label{omeg}
\omega = \int \! U \, dx + \int \! V \, dy - \int \partial_y \int \! U \, dx \,
dy.
\end{equation}
The integrals appearing above are what we call {\it partial integrals} (otherwise, equation (\ref{omeg}) would not be true). Basicaly, a partial integral is performed by considering all but the integration variable as constants. Therefore, if we also use that $U(x,y)$ and $V(x,y)$ are rational functions, one can use the result (\ref{IntFx}) thus obtaining:
\begin{equation}
\label{IntU}
\int \! U \, dx = u_0(x,y)+\sum_{i=1}^{K^u} C^u_i\, \ln(u_i(x,y))
\end{equation}
and
\begin{equation}
\label{IntV}
\int \! V \, dy = v_0(x,y)+\sum_{i=1}^{K^v} C^v_i\, \ln(v_i(x,y))
\end{equation}
where $u_i \,\, (i=0,...,K^u),\,\,v_i \,\, (i=0,...,K^v)$ are rational functions of $x,y$, and $C^u_i \,\,(i=1,...,K^u) , C^v_i \,\,(i=1,...,K^v)$ are constants.
We also have, using (\ref{IntU}):
\begin{equation}
\label{DuIntU}
\partial_y\int \! U \, dx = \partial_yu_0+\sum_{i=1}^{K^u} C^u_i\, \frac{\partial_yu_i}{u_i}.
\end{equation}
It is straightforward that the righthand side of the above equation is also a rational function on $(x,y)$. So, $\int \partial_y \int \! U \, dx \, dy$ can be written on a similar form as presented on eqs. (\ref{IntU},\ref{IntV}). Thus, using all these results, we can conclude that $w$ can be writen as:
\begin{equation}
\label{eqW}
w(x,y) = e^{w_0(x,y)+\sum_{i=1}^{K^w} C^w_i\, \ln(w_i(x,y))}.
\end{equation}
where $w_i \,\, (i=0,...,K^w)$ are rational functions of $x,y$, and $C^w_i \,\,(i=1,...,K^w)$ are constants.
So, combining theorem 1 and eq. (\ref{eqW}), it is straightforward to see that:
\begin{equation}
R = e^{(\int(Udx+Vdy))} = e^{(\int(d\omega (x,y)))} = e^{\omega (x,y)} = 
e^{w_0(x,y)+\sum_{i=1}^{K^w} C^w_i\, \ln(w_i(x,y))}
\end{equation}
\noindent
resulting in:
\begin{equation}
R = e^{w_0(x,y)} \prod_{i=1}^{K^w} w_i(x,y)^{C^w_i}
\end{equation}
\noindent
We can write each $w_i(x,y)$ as:
\begin{equation}
\label{wi}
w_i(x,y) = \prod_{j=1}^{n_i} p_j(x,y)^{k_j}
\end{equation}
\noindent
where the $p_j(x,y)$ are `monics' in $(x,y)$.
We can write, finally:
\begin{equation}
\label{ourIntFac}
R = e^{r_0(x,y)} \prod_{i=1}^{n} p_i(x,y)^{c_i}.
\end{equation}
\bigskip 
Notice that, if $r_0(x,y)$ is constant, the integrating factor $R$ would be
within reach of the PS-method. In what follows, we are going to present a method
to find $R$ for a class of equations where $r_0(x,y)$ is not a constant (so the usual
PS-method fails). 
\section{Finding the Integrating Factor}
\label{findingIF}
In this section, we introduce a method of finding the integrating factor for a class of LFOODEs using an algorithym somewhat similar to the PS-method. First we analyse the generality of our method. We then detail it and finally present some examples of its applicability.
\subsection{Introduction}
\label{subintro}
On the previous section, we have deduced the general form for the integrating factor for LFOODEs that can be written as in eq. (\ref{FOODE}).
By using $R$ as given on (\ref{ourIntFac}) onto eq. (\ref{eq_PS}), one finds:
\begin{equation}
\label{eqadendum}
D[r_0(x,y)] + \sum_i {\frac{c_iD[p_i]}{p_i}} = - \left( {\frac{\partial N}{\partial x}} + 
{\frac{\partial M}{\partial y}} \right),
\end{equation}
\bigskip
\bigskip
In the PS-method, the analogous to the above equation is eq. (\ref{eq_ni}):
$$
\sum_i {\frac{n_iD[f_i]}{f_i}} = -\left( {\frac{\partial N}{\partial 
x}} + {\frac{\partial M}{\partial y}} \right).
$$
Note that the main difference among these equations is the presence of the ``extra'' term $D[r_0]$ on eq. (\ref{eqadendum}). For eq. (\ref{eq_ni}), using the fact that $M$ and $N$ are polynomials on $(x,y)$, one can prove that $f_i$ are ``eigen-polynomials'' of the operator $D$, i.e., $f_i | D[f_i]$\cite{PS}. Regarding eq. (\ref{eqadendum}), if we had that $D[r_0]$ is a polynomial, by the same line of reasoning, we would conclude that $p_i$ are ``eigen-polynomials'' of the operator $D$, i.e., $p_i | D[p_i]$. 
\bigskip 
{\bf Conjecture 1:} {\it Consider a LFOODE of the form $dy/dx=M(x,y)/N(x,y)$, where $M(x,y)$
and $N(x,y)$ are polynomials in $(x,y)$, with integrating factor given by
$R = e^{r_0(x,y)} \prod_{i=1}^{n} p_i(x,y)^{c_i}$, where $r_0$ is a rational function on $(x,y)$, the $p_i$'s are irreducible polynomials on $(x,y)$ and the $c_i$'s are constants. Then we have that $D[r_0]$ is a polynomial ($D \equiv N \partial_x+ M \partial_y$).}
\bigskip
Assuming this to be true, we are going to present a method to find the integrating factor for a class of LFOODEs missed by the PS-method.
\subsection{The Method}
\label{mmethod}
The method we are going to introduce 1now deals with three different general forms for $r_0(x,y)$:
\begin{enumerate}
\item $r_0 = r_0(x)$
\item $r_0 = r_0(y)$
\item $r_0 = r(x)+s(y)$
\end{enumerate}
where all of the above functions are rationals.
The LFOODEs that correspond to these cases define the class of equations solved by our method.
\subsubsection{Case $r_0=r_0(x)$}
\label{case1}
Reminding the reader that the $D$ operator is defined by 
$D \equiv N \partial_x+ M \partial_y$, eq. (\ref{eqadendum}) will then become:
\begin{equation}
\label{eqadendum2}
N \frac{dr_0(x)}{dx} + \sum_i {\frac{c_iD[p_i]}{p_i}} = - \left( {\frac{\partial N}{\partial x}} + 
{\frac{\partial M}{\partial y}} \right).
\end{equation}
First of all, since we made a hypothesis implying that the $p_i$'s are ``eigenpolynomials'' of the $D$ operator, we have: $D[p_i] = g_i\,\, p_i$, where $g_i$ are polynomials called eigenvalues of $D$. We can then calculate all the $p_i$ and associated $g_i$ up to a given order, in the same fashion of the PS-method.
We can then write eq.(\ref{eqadendum2}) as:
\begin{equation}
\label{eqadendum3}
N \frac{dr_0(x)}{dx} = - \left( {\frac{\partial N}{\partial x}} + 
{\frac{\partial M}{\partial y}} \right) - \sum_i {c_i\,\, g_i}.
\end{equation}
leading to:
\begin{equation}
\label{r0x}
\frac{dr_0(x)}{dx} = - \frac{N_x + M_y + \sum_i {c_i\,\, g_i}}{N}.
\end{equation}
In order to find $r_0(x)$ we need first to integrate eq.(\ref{r0x}):
\begin{equation}
\label{r0}
r_0(x) = - \int{\frac{N_x + M_y + \sum_i {c_i\,\, g_i}}{N}}\,{dx} + C.
\end{equation}
Since $r_0$ is a {\it rational} function of $x$ only, we can determine the $c_i$'s by imposing that the derivative of the righthand side of eq.(\ref{r0}) with respect to $y$ be equal to zero and that the logarithmic terms that might arise from the integration vanish:
If we manage to find a set of $c_i$'s satisfying these conditions, in the order we are considering, we will have found the integrating factor for the LFOODE. Otherwise, we then increase the order of the $p_i$'s and try again, until we succeed, in the same manner as in the PS-method.
\subsubsection{Case $r_0=r_0(y)$}
Analogously, for this case, eq.(\ref{eqadendum}) will become:
\begin{equation}
\label{eqadendum4}
M \frac{dr_0(y)}{dy} + \sum_i {\frac{c_iD[p_i]}{p_i}} = - \left( {\frac{\partial N}{\partial x}} + 
{\frac{\partial M}{\partial y}} \right).
\end{equation}
Following a procedure similar to the one used on section \ref{case1}, we can write $r_0(y)$ as:
\begin{equation}
\label{r02}
r_0(y) = - \int{\frac{N_x + M_y + \sum_i {c_i\,\, g_i}}{M}}\,{dy} + C.
\end{equation}
Since $r_0$ is a {\it rational} function of $y$ only, we can determine the $c_i$'s by imposing that the derivative of the righthand side of eq.(\ref{r02}) with respect to $x$ be equal to zero and that the logarithmic terms that might arise from the integration vanish:
In the same way, if we manage to find a set of $c_i$'s satisfying these conditions, in the order we are considering, we will have found the integrating factor for the LFOODE. Otherwise, we then increase the order of the $p_i$'s and try again, until we succeed, in the same manner as in the PS-method.
\subsubsection{Case $r_0=r(x) + s(y)$}
For this case, eq.(\ref{eqadendum}) becomes:
\begin{equation}
\label{r+s1}
N \frac{dr(x)}{dx} + M \frac{ds(y)}{dy} +\sum_i {\frac{c_iD[p_i]}{p_i}} = - \left( {\frac{\partial N}{\partial x}} + 
{\frac{\partial M}{\partial y}} \right).
\end{equation}
Dividing the equation above by $N$ and isolating $\frac{dr(x)}{dx}$, one obtains:
\begin{equation}
\label{rx}
\frac{dr(x)}{dx} = -\frac{M}{N} \frac{ds(y)}{dy} - \frac{N_x + M_y + \sum_i {c_i\,\, g_i}}{N}.
\end{equation}
Integrating both sides of the equation, with respect to $x$:
\begin{equation}
\label{r}
r(x) = - \frac{ds(y)}{dy} \int{\frac{M}{N}}\,{dx} - \int{\frac{N_x + M_y + \sum_i {c_i\,\, g_i}}{N}}\,{dx}.
\end{equation}
We can then solve (\ref{r}) to find:
\begin{equation}
\label{sy}
\frac{ds(y)}{dy} = \frac{\displaystyle{\int{\frac{N_x + M_y + \sum_i {c_i\,\, g_i}}{N}}\,{dx} - r(x)}}{\displaystyle{\int{\frac{M}{N}}\,{dx}}}.
\end{equation}
Analogous procedure leads to:
\begin{equation}
\label{rxfinal}
\frac{dr(x)}{dx} = \frac{\displaystyle{\int{\frac{N_x + M_y + \sum_i {c_i\,\, g_i}}{M}}\,{dy} - s(y)}}{\displaystyle{\int{\frac{N}{M}}\,{dy}}}.
\end{equation}
The $c_i$'s remain to be determined. Once again, we still have to imposed that, by construction, $r(x)$ and $dr(x)/dx$ do not depend on $y$ and, analogously, $s(y)$ and $ds(y)/dy$ do not depend on $x$. By imposing that, and that both $r(x)$ and $s(y)$ are {\it rational} functions, we may hope to find a set of suitable values for $c_i$. If that is the case, we would have found $r(x)$ and $s(y)$ and, consequently, the integrating factor. If this fails, we have to increase the order of the $p_i$'s and try all over again.
\subsection{Examples}
In this section, we are going to present one example of application of our method for each of the cases mentioned above.
\subsubsection{$r_0=r_0(x)$}
\label{case1}
Consider the LFOODE (eq.{\bf I.129} in \cite{kamke}):
\begin{equation}
\left (x+1\right ){\frac {d}{dx}}y(x)+y(x)\left (y(x)-x\right ).
\end{equation}

For this equation, up to order 1, we have that the ``eigenpolynomials'' (with the associated ``eigenvalues'') are:
\begin{itemize}
\item $p_1 = y,\,\,\,\,\,g_1 = (x-y)$, 
\item $p_2 = (x+1),\,\,\,\,\,g_2 = 1$. 
\end{itemize}
So, eq.(\ref{r0}) becomes:
\begin{equation}
\left (\left (2+{\it c_1}\right )y-{\it c_2}+{\it c_1}\right )\ln (x+1)-x
-{\it c_1}\,x.
\end{equation}
Leading to:
\begin{equation}
c_1 = -2,\,\,\,\,\,c_2 = -2,
\end{equation}
and, consequently to:
\begin{equation}
r_0(x) = x,\,\,\,\,\,\,\rightarrow\,\,\,\,\,R = \frac{e^x}{y^2(x+1)^2} 
\end{equation}

\subsubsection{$r_0=r_0(y)$}
\label{case2}
Consider the LFOODE (eq.{\bf I.235} in \cite{kamke}):
\begin{equation}
\left (x+1\right ){\frac {d}{dx}}y(x)+y(x)\left (y(x)-x\right ).
\end{equation}

For this equation, up to order 1, we have that the ``eigenpolynomials'' (with the associated ``eigenvalues'') are:
\begin{itemize}
\item $p_1 = y,\,\,\,\,\,g_1 = -b$, 
\end{itemize}
So, eq.(\ref{r0}) becomes:
\begin{equation}
\left (-1-{\it c_1}\right )\ln (y)+{\frac {y}{b}}\end{equation}
Leading to:
\begin{equation}
c_1 = -1,
\end{equation}
and, consequently to:
\begin{equation}
r_0(y) = \frac{y}{b},\,\,\,\,\,\,\rightarrow\,\,\,\,\,R = \frac{e^{\frac{y}{b}}}{y} 
\end{equation}

\subsubsection{$r_0=r(x)+s(y)$}
\label{case3}
Consider the Abel LFOODE of the first kind:
\begin{equation}
{\frac {d}{dx}}y(x)={\frac {y(x)^{2}\left (y(x)+x-1
\right )}{{x}^{2}}}.
\end{equation}

For this equation, up to order 1, we have that the ``eigenpolynomials'' (with the associated ``eigenvalues'') are:
\begin{itemize}
\item $p_1 = x,\,\,\,\,\,g_1 = x$,
\item $p_2 = y,\,\,\,\,\,g_2 = y^2+y*x-y$, 
\item $p_3 = x+y,\,\,\,\,\,g_3 = x-y+y^2$. 
\end{itemize}
So, from eq.(\ref{rxfinal}) we get:
\begin{eqnarray}
\frac{dr(x)}{dx} = -{\frac {\left (\left (-{\it c_2}-2\right ){x}^{2}+\left (2\,{\it c_3}+2\,{\it c_2
}+6+{\it c_1}\right )x-{\it c_3}-{\it c_2}-2\right )y\ln (y
)}{{x}^{2}\left (x-1+\ln (y
)y-\ln (y+x-1)y\right )}}+ \hspace{1ex}\nonumber\\
\nonumber\\
-{\frac {\left (\left (-{\it c_3}-1\right ){x}^{2}-{\it c_1}\,x-1\right )y\ln (y+x-1
)}{{x}^{2}\left (x-1+\ln (y
)y-\ln (y+x-1)y\right )}}+\hspace{32ex} \nonumber \\
\nonumber\\
-{\frac {\left (-{x}^{2}+2\,x-1\right )ys(y)+\left ({\it c_3}+2+{\it c_1}\right )
{x}^{2}+\left (-{\it c_1}-{\it c_3}-2\right )x
}{{x}^{2}\left (x-1+\ln (y
)y-\ln (y+x-1)y\right )}} = 0\hspace{10ex}
\end{eqnarray}
Leading to:
\begin{eqnarray}
\left (-{\it \frac{dr(x)}{dx}}\,{x}^{2}y+\left (\left (-{\it c_3}-1\right ){x}^{2}-{
\it c_1}\,x-1\right )y\right )\ln (y+x-1)+ \hspace{17ex}\nonumber\\
\left ({\it \frac{dr(x)}{dx}}{x}^{2}y+
\left (\left (-{\it c_2}-2\right ){x}^{2}+\left (2\,{\it c_3}+2{\it c_2
}+6+{\it c_1}\right )x-{\it c_3}-{\it c_2}-2\right )y\right )\ln (y)+ \nonumber\\
\left ({x}^{3}-{x}^{2}\right ){\it \frac{dr(x)}{dx}}+\left (-{x}^{2}+2\,x-1\right )y
s(y)+ \hspace{36ex}\nonumber\\
\label{rxcomlogs}
\left ({\it c_3}+2+{\it c_1}\right ){x}^{2}+\left (-{\it c_1}-{\it 
c_3}-2\right )x = 0 \hspace{39ex}
\end{eqnarray}
Since $dr(x)/dx$ and $s(y)$ are rational functions, in the above equation the coefficients of the logarithmic terms must vanish. Therefore:
\begin{eqnarray}
\label{rxexemplo1}
- \frac{dr(x)}{dx}\,{x}^{2}y+\left (\left (-{\it c_3}-1\right ){x}^{2}-{\it c_1}
\,x-1\right )y =0 \hspace{30ex}\\
\label{rxexemplo2}
\frac{dr(x)}{dx}\,{x}^{2}y+\left (\left (-{\it c_2}-2\right ){x}^{2}+\left (2\,
{\it c_3}+2\,{\it c_2}+6+{\it c_1}\right )x-{\it c_3}-{\it c_2}-2\right )y = 0 \hspace{1ex}
\end{eqnarray}
Solving (\ref{rxexemplo1}) and (\ref{rxexemplo1}) for $dr(x)/dx$ and integrating in $x$, one gets: 
\begin{equation}
\label{rr1}
r(x)=-x-{\it c3}\,x+{x}^{-1}-{\it c1}\,\ln (x))
\end{equation}
\begin{equation}
\label{rr2}
r(x)=\left (-2\,{\it c3}-2\,{\it c2}-6-{\it c1}\right )\ln (x)+{\it c2
}\,x+2\,x-2\,{x}^{-1}-{\frac {{\it c2}}{x}}-{\frac {{\it c3}}{x}}
\end{equation}
Imposing that the logarithmic terms vanish and that the right-hand side of (\ref{rr1}) and (\ref{rr2}) are equal, we get:
\begin{equation}
\label{cs}
\left \{{\it c1}=0,{\it c2}=-{\it c3}-3,{\it c3}={\it c3}\right \}.
\end{equation}
Using these in (\ref{r+s1}), solving the resulting equation for $ds(y)/dy$ and imposing that it can not deppend on $x$, we have: 
\begin{equation}
\left (1+{\it c3}\right ){x}^{2}+\left (\left (2+2\,{\it c3}\right )y-
2-2\,{\it c3}\right )x+\left (1+{\it c3}\right ){y}^{2}+\left (-2-2\,{
\it c3}\right )y+1+{\it c3}=0,
\end{equation}
implying that $c_3 = -1$.
So, using (\ref{cs}):
\begin{equation}
\label{csfinal}
\left \{{\it c1}=0,{\it c2}=-2,{\it c3}=-1\right \}.
\end{equation}
Substituting (\ref{csfinal}) into (\ref{rr1}), we get $r(x)=1/x$. Using this in eq.(\ref{rxcomlogs}) and solving for $s(y)$, we have finally:
\begin{equation}
r(x) = \frac{1}{x},\,\,\, s(y) = \frac{1}{y},\,\,\,\rightarrow\,\,\,\,R = \frac{e^{\frac{1}{x}+\frac{1}{y}}}{y^2(x+y)}.
\end{equation}
\section{Conclusion}
In this paper, we have presented an extension of the Prelle-Singer method which solves a class of LFOODEs missed by the PS-approach. Although our method deals with a restricted class of LFOODEs (of the type (\ref{FOODE}) and among those the ones defined by the three cases considered on section \ref{mmethod}) and we are still working on the demonstration of {\bf conjecture 1} (see section \ref{subintro}) that would prove the generality of the method, we believe the method to be a valid contribuition since we could not find any counterexample where the conjecture fails and, apart from that, it solves some LFOODEs that ``escape'' from many powerfull solvers, using any method of solution (the example in section \ref{case3} is such a case).
We hope, in the near future, to present results concerning the extension of our method to deal with a larger class of LFOODEs. For example, considering $r_0(x,y)$ to be a general rational function or LFOODEs with $M(x,y)$ and/or $N(x,y)$ not restricted to be polynomials, etc.

\end{document}